\documentclass[10pt]{iopart}
\usepackage{graphicx}
\usepackage{bm}
\usepackage{iopams}
\usepackage{bbm}
\usepackage{mathrsfs} 
\usepackage[usenames]{color}

\newcommand{\image}[4]{
\begin{figure}
   \begin{center}
      \includegraphics[angle = 0, width = #2]{#1}
   \end{center}
\caption{\label{#4}#3}
\end{figure}
}
\newcommand{\bra}[1]{\langle #1|}
\newcommand{\ket}[1]{| #1 \rangle}
\newcommand{\braket}[2]{\langle #1|#2 \rangle}
\newcommand{\expi}[1]{ {\rm e}^{{\rm i} #1}}
\newcommand{\expim}[1]{ {\rm e}^{-{\rm i} #1}}
\newcommand{\cre}[1]{\hat{#1}^\dagger}  
\newcommand{\des}[1]{\hat{#1}}  

\newcommand{\op}[1]{\hat{#1}} 

\newcommand{\eqref}[1]{(\ref{#1})} 

\begin{document}
\title{Fast initialization of a high-fidelity quantum register using optical superlattices.}

\author{B.~Vaucher, S.R.~Clark, U.~Dorner, and D.~Jaksch}


\address{Clarendon Laboratory, University of Oxford, Parks Road, OX1 3PU,
United Kingdom}

\ead{benoit.vaucher@merton.ox.ac.uk}
\submitto{\NJP}

 \begin{abstract}
We propose a method for the fast generation of a quantum register of
addressable qubits consisting of ultracold atoms stored in an
optical lattice. Starting with a half filled lattice we remove every
second lattice barrier by adiabatically switching on a superlattice
potential which leads to a long wavelength lattice in the Mott
insulator state with unit filling. The larger periodicity of the
resulting lattice could make individual addressing of the atoms via
an external laser feasible. We develop a Bose-Hubbard-like model for describing the dynamics of cold atoms in a lattice when doubling the lattice periodicity via the addition of a superlattice potential. The dynamics of the transition from a
half filled to a commensurately filled lattice is analyzed
numerically with the help of the Time Evolving Block Decimation
algorithm and analytically using the Kibble-Zurek theory. We show
that the time scale for the whole process, i.e. creating the half
filled lattice and subsequent doubling of the lattice periodicity, is
significantly faster than adiabatic direct quantum freezing of a
superfluid into a Mott insulator for large lattice periods. Our
method therefore provides a high fidelity quantum register of
addressable qubits on a fast time scale.
\end{abstract}

\maketitle

\section{Introduction}
\image{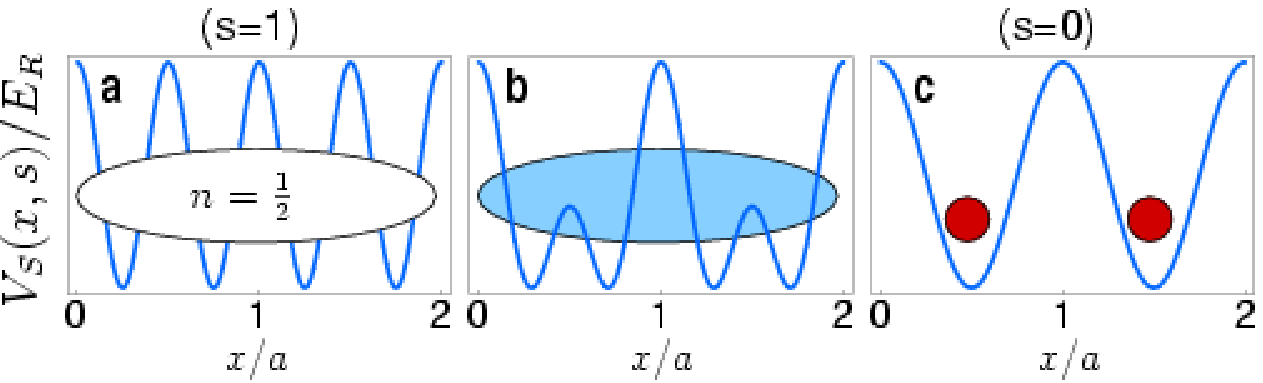}{10cm}{(a) The initial profile of the lattice
is associated with the value of  $s=1$. The periodicity of the
lattice is then progressively doubled (b) until it reaches its final
profile (c) associated with the parameter $s=0$. The number of sites
$M$ corresponds to the number of unit cells in the large lattice
limit. By starting with a filling factor of $n=1/2$, this procedure leads to a lattice with filling factor of $n=1$.}{profile}


Systems of cold atoms trapped in optical lattices provide the unique
opportunity to coherently manipulate a large number of
atoms~\cite{Jaksch2005,Lewenstein2006,Cirac2004}. The remarkable
degree of experimental control offered by these systems, as well as
the possibility to use the internal hyperfine states of the atoms to
encode qubits, make them particularly suited for quantum information
processing (QIP).  In this context, optical lattices in a
Mott-insulating (MI) state with unit filling can be viewed as the
realization of a quantum register, and it is possible to
collectively manipulate the qubits stored in such a register
experimentally~\cite{Jaksch1999,Mandel2003}. However, in many
quantum computing schemes based on neutral atoms stored in optical
lattices the application of single qubit gates~\cite{Cirac2004} or
single qubit measurements~\cite{Raussendorf2001} requires the
ability to address single atoms with a focused laser beam. This
remains experimentally challenging since these operations have to be
performed without perturbing the state of other atoms in their
vicinity.

A number of strategies have been proposed to circumvent this problem
by using global operations \cite{Vollbrecht2004,Zhou2006,Kay2004},
e.g. via ``marker atoms'' which are moved to a particular
lattice site and interact with the corresponding register qubit such
that an external laser affects only that qubit \cite{Calarco2004}.
Another way is simply to use a quantum register in which the atoms
are distant enough such that they can be addressed individually by a
laser. This method requires an optical lattice with filling factor
$n=1$ in the MI state with a sufficiently large distance between
the atoms~\cite{Haffner2005}. The
initialization time of a MI state is proportional to the tunneling time of the atoms between neighboring sites. Therefore, by using the conventional method of quantum-freezing a superfluid (SF) state
\cite{Jaksch1999,Greiner2002b,Julienne2005,Sklarz2002}, it scales exponentially with the lattice
spacing \cite{Clark2004,Blakie2004,Rey2004}.

In this paper we propose an alternative method to generate a long
wavelength lattice with one atom per site in the MI state. Starting
with a one dimensional lattice with a short period---and hence a short initialization time---and filling factor $n=1/2$ we remove
every second potential barrier by adiabatically turning on a
superlattice. This superlattice potential has already been
experimentally realized~\cite{Peil2003,Sebby-Strabley2006}. This
procedure eventually leads to a long wavelength lattice in which the periodicity has been doubled and where the atoms
are in a MI state  with $n=1$ (see figure~\ref{profile}). This scheme does not require changing the
angles of the intersecting laser beams. Furthermore, we show that our method allows for the
initialization of a MI state with unit filling factor on a
time-scale which, although scaling exponentially with the final lattice spacing, is approximately one order of magnitude smaller than the direct quantum-freezing method. Although we only consider the case of an homogeneous lattice, the results presented in this paper extend to the case of weak harmonic confinements, quartic~\cite{Gygi2006} and box traps~\cite{Meyrath2005}.

This paper is structured as follows: In Sec.~\ref{model} we
introduce the model used to describe the system dynamics. In
Sec.~\ref{numerics} we discuss ground state properties, particularly
two-site correlation functions and quasi momentum distributions.
Also, we present and discuss numerical results for the probability
of staying in the ground state during the transition depending on
the speed of the ramping. In Sec.~\ref{analytics} we apply the
analytical Kibble-Zurek theory and compare it with our numerical
results. Finally, we summarize and conclude in
Sec.~\ref{conclusion}.

\section{Model}
\label{model}

We consider a gas of interacting ultracold bosonic atoms
loaded into a three dimensional optical lattice. The lattice is
formed by pairwise orthogonal standing wave laser fields and its
optical potential is given by~\cite{jaksch1}
\begin{equation}
V_{OL}({\bf r},s)=V_S(x,s) + V_T\,[\sin^2(k z)+\sin^2(k y)].
\end{equation}
Here $V_T$ is the depth of the potential in the $y$-- and
$z$--directions created by pairs of lasers with wave number $k=2
\pi/\lambda$, wave length $\lambda$ and period $a_T=\lambda/2$. The extension to the case of
different optical potentials in the $y$-- and $z$--directions is
straightforward. In the $x$--direction two pairs of laser beams with a long wavelength $\lambda_L$ and short wavelength $\lambda_S=\lambda_L/2$ are applied. The potential
in the $x$-direction is thus given by
\begin{equation}
V_S(x,s)= V_0\,(1-s)\, \sin^2\left(k_L x\right) + V_0\,s\,
\sin^2\left(k_S x\right),
\end{equation}
with $V_0$ the depth of the lattice, $k_L=2 \pi/\lambda_L$ and $k_S=2 \pi/\lambda_S$. The
depths of the potentials will be expressed in units of the recoil
energy $E_R=k_L^2/2m$ with $m$ the mass of the atoms (taking
$\hbar=1$ throughout). The parameter $s \in [0,1]$ is determined by
the relative intensities of the two pairs of lasers. By changing $s$
from $1$ to $0$ the lattice profile is continuously transformed from
a sinusoidal potential with a small period $a_S=\lambda_S/2$ to one with a long period
$a=\lambda_L/2$, thus halving the number of lattice sites per unit
length along the $x$-direction (see~figure~\ref{profile}). The
lattice constant $a$ corresponds to the size of a unit cell for 
$s<1$~\footnote{To avoid any discontinuity, we work with a lattice periodicity of $a$ even in
the case $s=1$.}. We refer to the lattice profile with parameters
$s=1$ and $s=0$ as to the {\it small lattice limit} and the {\it large lattice limit}, respectively.

The Hamiltonian of the system in second quantization reads
\begin{equation}
\hat{H}=\int {\rm d}{\bf r} \, \op{\psi}^{\dag}({\bf r})\, \op{h}_0(\bar{\bf r})\, \op{\psi}({\bf r})
+ \frac{g}{2}\int {\rm d}{\bf r}\, \op{\psi}^\dag({\bf r})\,
\op{\psi}^\dag({\bf r})\, \op{\psi}({\bf r})\, \op{\psi}({\bf
r}),\label{fullhamilt}
\end{equation}
where $\op{h}_0(\bar{\bf r})=-(1/{2 m }) \nabla^2 + V_{OL}(\bar{\bf
r})$ is the one-particle Hamiltonian. The symbol $\bar{\bf r}=({\bf
r},s)$ represents the position variable ${\bf r}$ and lattice
parameter $s$. The interaction between the atoms is modelled by
$s$-wave scattering with $g=4 \pi a_s /m$ where $a_s$ is the $s$-wave
scattering length. The bosonic field operators obey the usual
commutation relations $[\psi({\bf r}),\psi^{\dag}({\bf
r'})]=\delta({\bf r}-{\bf r'})$ with $\delta$ denoting the Dirac
delta function.

We restrict our considerations to the case where $V_T$ is
sufficiently large so that motion along the $y$-- and
$z$--directions is frozen. The dynamics of the system is then effectively one dimensional along the $x$--direction. As shown in figure~\ref{bandstructure}a the two lowest bands of the Hamiltonian $\op{h}_{0,x}=-(1/2 m)\left({{\rm d}}/{{\rm d}x}\right)^2 +
V_{S}(x,s)$ are separated in energy
by much less than the typical motional excitation energy $E_{\rm ex}
= \sqrt{4 V_0 E_R}$~\cite{jaksch1} for values $s \approx 1$
\footnote{The two lowest bands form two segments of the lowest Bloch band if a cell
size of $a/2$ is used in the case $s=1$.}. Therefore, despite
assuming that atoms loaded into the lattice are ultracold, we have
to consider the two lowest Bloch bands in $x$--direction to obtain an
accurate description of the atomic dynamics. However, excitations to
higher bands in the $y$-- and $z$--directions are neglected in our
investigations since we assume that the temperature of the atomic cloud is $k_B T \ll
E_{\rm ex}$. We then expand the bosonic field operator as
\begin{equation}
 \op{\psi}(\bar{\bf r})=\sum_{i=1}^M \, \phi_{a,i}(\bar{\bf r})\,
 \des{a}_i + \sum_{i=1}^M \, \phi_{b,i}(\bar{\bf r})\,\des{b}_i, \label{fieldop}
\end{equation}
where $M$ is the number of lattice sites and $\cre{\alpha}_i$
($\alpha=a,b$) creates a particle in the mode associated with the
localized function $\phi_{\alpha,i}(\bar{\bf r})$ centered at site
$i$. The mode functions $\phi_{\alpha,i}(\bar{\bf
r})=w_{\alpha,i}(x,s) W_{i,0}(y)W_{i,0}(z)$ are factorized into a
product of well localized Wannier functions (WF) $W_{i,0}$ of the
lowest Bloch band in the $y$-- and $z$--directions
\cite{Ashcroft1976} and mode functions $w_{\alpha,i}(x,s)$ in
$x$--direction.

The aim of the next section is to describe the single particle
dynamics in the tight binding (TB) approximation. If we were to use
Wannier functions for $w_{\alpha,i}(x,s)$ this approximation would
restrict our model to sinusoidal Bloch bands~\cite{Schnell2002}.
Because of the deviation of the lowest two bands from a sinusoidal
dispersion relation (see figure~\ref{bandstructure}a) when $s\approx
1$ we instead use generalized Wannier function (GWFs) for
$w_{\alpha,i}(x,s)$ (see \ref{gwf} for a detailed definition and a
description of their properties)~\cite{Marzari1997}. By exploiting
the optimization procedure described in \ref{gwf} we calculate GWFs
$w_{\alpha,i}(x,s)$ which are well localized at lattice sites $i$.
Typical shapes of GWFs and the effects of optimizing their
localization are shown in figure~\ref{wannier}. We note that these
GWFs are in general composed of superpositions of Bloch orbitals of
both bands and are not related to Wannier functions by a local
transformation. Only when $s\approx 0$ the $w_{\alpha,i}(x,s)$ are
equivalent to the Wannier functions of the first and the second
Bloch band, respectively. Finally, they are always~(anti-)symmetric
with respect to the center of the lattice site $i$ for $\alpha = a$
($\alpha=b$).

\image{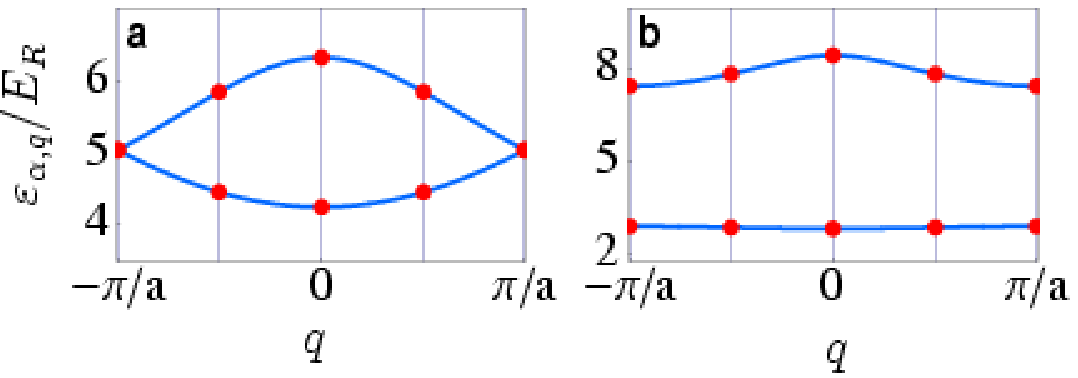}{10cm}{Band structure along the $x$--direction in (a) the small and (b) large lattice
limit for $V_0=10 E_R$. The points represents the values of q for
$q=0,\,\pi/a,\,\pi/2a$. The value of $\varepsilon_{\alpha,q}$ correspond to the energy of
single-particles with momentum $q$ in the $\alpha$--th Bloch band.
In the small lattice limit, the two first Bloch bands are
connected.}{bandstructure}

\subsection{Single-particle Hamiltonian}
\label{onepart}

\image{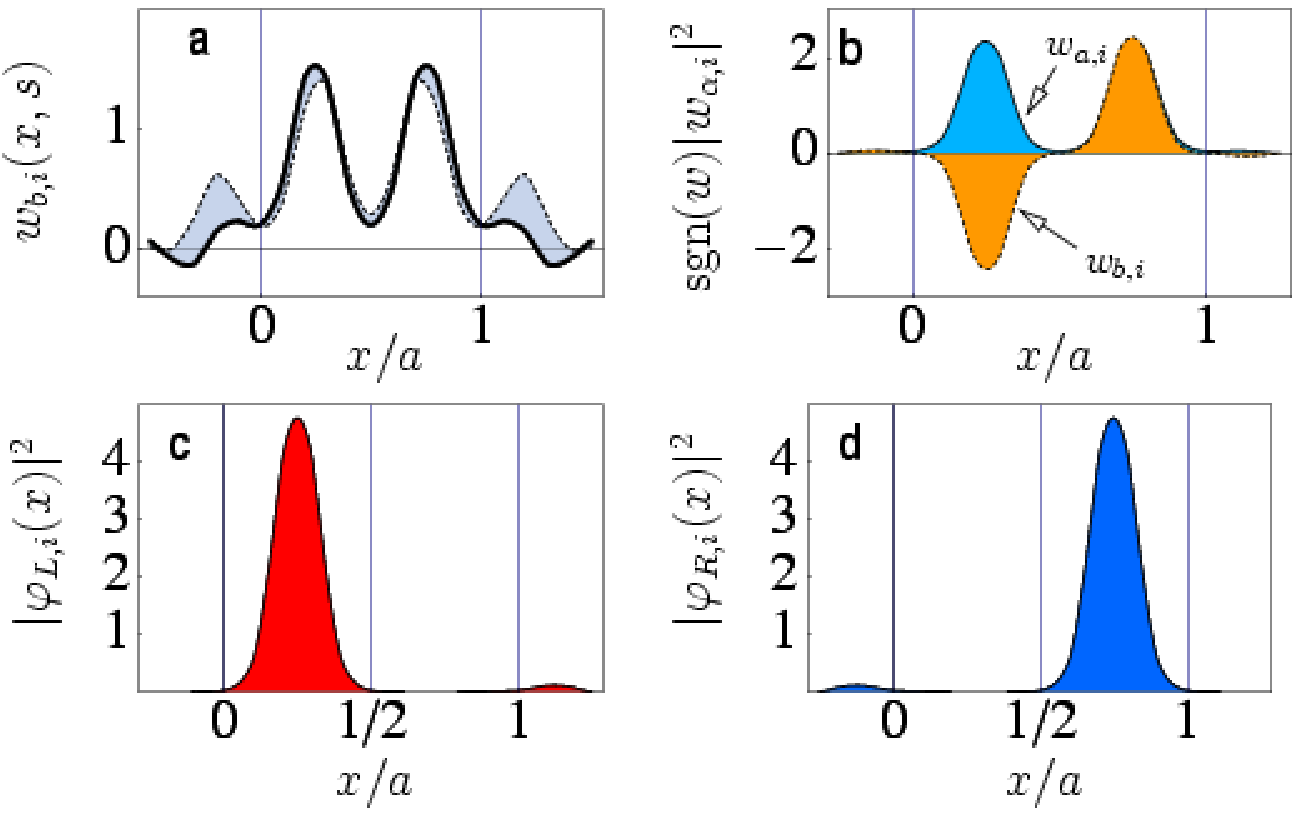}{10cm}{(a) Optimized (thick line) and
non-optimized (dashed line) GWFs associated with the first mode for
$V_0=30 E_R$ and the lattice profiles $s=1$. The area between the
optimized and non-optimized mode functions is shaded in order to
illustrate how the localization procedure reduces the spread of the
optimized function. (b) The square of the mode functions for $V_0=30
E_R$ and $s=1$. By combining the mode functions shown in (b), we can construct two new mode functions corresponding to
particles localized in either the left (c) or the right (d) well
of a given site.}{wannier}

Inserting the approximate field operator equation~\eqref{fieldop} into
the first term of equation~\eqref{fullhamilt} yields the single-particle
part of the Hamiltonian in terms of $\cre{a}_i$ and  $\cre{b}_i$.
Applying the tight binding approximation, which amounts to keeping
only nearest-neighbor hopping terms, the single-particle Hamiltonian
can be approximated by
\begin{eqnarray}
\hat{H}_0(s) &=\sum_{i=1}^{M-1}\left(   J_{bb}(s)\, \cre{b}_i \des{b}_{i+1} -J_{aa}(s)\, \cre{a}_i \des{a}_{i+1}+ {\rm h.c}\right)\nonumber\\
&+ \sum_{i=1}^{M-1}\left( J_{ba}(s)\, \cre{b}_i \des{a}_{i+1}-J_{ab}(s)\, \cre{a}_i \des{b}_{i+1} + {\rm h.c.}\right) \nonumber\\
&+  \sum_{i=1}^{M}\left( V_{a}(s)\, \cre{a}_i \des{a}_{i} + V_{b}(s)\, \cre{b}_i \label{oneparthamilt}
\des{b}_{i}\right),
\end{eqnarray}
where
\begin{equation}
J_{\alpha \beta}(s)= \int {\rm d}x\, w^*_{\alpha,i}(x,s)
\,\op{h}_{0,x}(s)\,w_{\beta,i+1}(x,s),\label{paramJgen}
\end{equation}
is the hopping matrix element between neighboring sites along the
$x$--axis and
\begin{equation}
V_{\alpha} (s)= \int {\rm d}x\, w^*_{\alpha,i}(x,s)
\,\op{h}_{0,x}(s)\,w_{\alpha,i}(x,s),\label{paramEgen}
\end{equation}
is the local on-site energy of a particle in mode $\alpha$. Note that
hopping between modes $a$ and $b$ within one site is not allowed by
the symmetry properties of the GWFs. However, the inclusion of
non-zero hopping matrix elements $J_{ab}$ and $J_{ba}$ is essential
to accurately reproduce the single particle behaviour of the full
Hamiltonian~\eqref{fullhamilt}. The symmetry properties of WFs would
not allow the inclusion of these terms~\cite{Schnell2002}.

For periodic boundary conditions, the parameters $V_\alpha$ and
$J_{\alpha,\beta}$ can be found from the band structure without
explicit calculation of the mode functions (see \ref{paramEq}). The
numerical values of $V_\alpha$ and $J_{\alpha,\beta}$ for
$V_0=30E_R$ are shown in figure \ref{parameters}a. Using these
parameters, we find that $\op{H}_0(s)$ very accurately reproduces
the band structure of the exact Hamiltonian for all values of $s$,
thus justifying the utilization of the TB approximation and the
corresponding GWFs.

\image{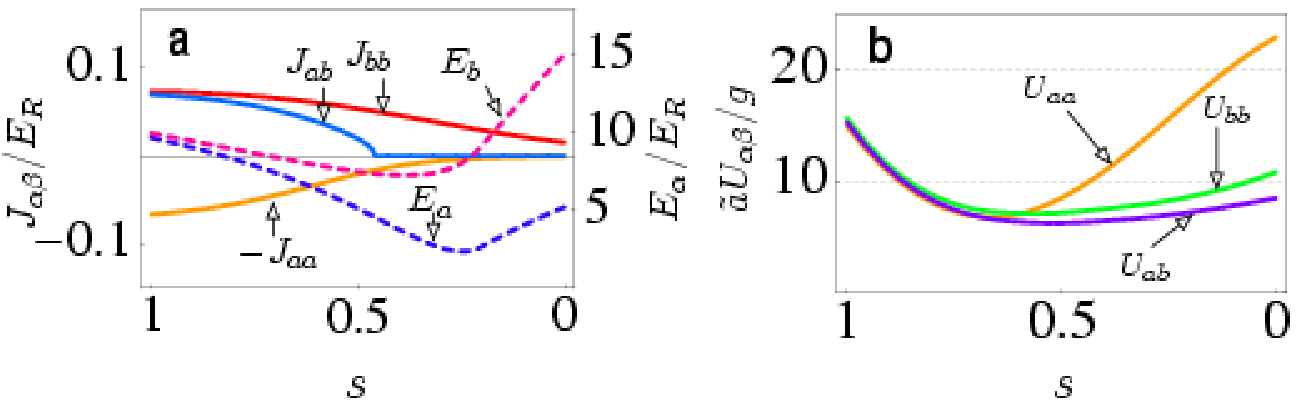}{11cm}{Parameters of the effective
Hamiltonian $\op{H}_{\rm eff}$ as functions of $s$ for a lattice depth
of $V_0=30 E_R$ and $V_T=60 E_R$ with $\tilde{a}=a\, a_T^2$. The hopping matrix elements (a) have been calculated using the method described in~\ref{paramEq}, while the on-site interaction energies (b) are
calculated using optimized GWFs.}{parameters}

\subsection{Interaction Hamiltonian}
\label{ihamilt}

To calculate the interaction matrix elements of $\op{H}$
the explicit form of the localized GWFs is needed. We find that for
$V_0 > 10 E_R$, off-site interaction terms are at least two orders
of magnitude smaller than on-site interactions. We therefore only keep
the dominant on-site terms and find the interaction Hamiltonian
$\op{H}_I$
\begin{eqnarray}
\hat{H}_I(s) &= \sum_{i=1}^{M}\frac{ U_{aa}(s)}{2}\, \des{n}^a_i\left(  \des{n}^a_i-1\right) + \frac{ U_{bb}(s)}{2}\, \des{n}^b_i\left(  \des{n}^b_i-1\right)\nonumber\\
&+ \sum_{i=1}^{M} \frac{U_{ab}(s)}{2}\,\left(4 \des{n}^a_i\des{n}^b_{i}+ \cre{b}_i\cre{b}_{i}\des{a}_i\des{a}_{i}+\cre{a}_i\cre{a}_{i}\des{b}_i\des{b}_{i}\right),\label{inthamilt}
\end{eqnarray}
where $\des{n}^a_i=\cre{a}_i \des{a}_{i}$ and $\des{n}^b_i=\cre{b}_i
\des{b}_{i}$ are the site-occupation number operators. The on-site
interaction matrix elements are given by
\begin{equation}
U_{\alpha,\beta}(s)= g \, \int d{\bf r}\, w^*_{\alpha,i}(\bar{\bf
r}) \, w^*_{\beta,i}(\bar{\bf r}) \,w_{\alpha,i}(\bar{\bf
r})\,w_{\beta,i}(\bar{\bf r}).
\end{equation}
The numerical values of $U_{\alpha,\beta}$ as a function of the
lattice profile $s$ are shown in figure~\ref{parameters}b. Figure
\ref{parameters}b shows that their values become equal as $s
\rightarrow 1$ for sufficiently large $V_0$.

Combining the single- and two-particle contributions the effective
Hamiltonian describing the system dynamics is given by
\begin{equation}
\op{H}_{\rm eff}(s)= \op{H}_0(s)+\op{H}_I(s)\label{finalh}.
\end{equation}
By using $\op{H}_{\rm eff}(s)$ for $s$ varying in time we implicitly
assume that the system adiabatically follows changes in the mode
functions $w_{\alpha,i}$. For all dynamical calculations carried out
in this work we have carefully chosen the time dependence of $s$ so
that such non-adiabatic contributions can safely be neglected.

\subsection{Limiting cases}\label{modeconv}

In the small lattice limit ($s=1$), the superpositions
$\varphi_{L,i}=(w_{a,i}-w_{b,i})/\sqrt{2}$ and
$\varphi_{R,i}=(w_{a,i}+w_{b,i})/\sqrt{2}$ correspond to mode
functions localized in the left and in the right well of site $i$
respectively (see figure~\ref{wannier}c and \ref{wannier}d). The
associated bosonic operators are defined by
\begin{equation}
\cre{c}_{L,i}=\frac{1}{\sqrt{2}} (\cre{a}_{i}-\cre{b}_{i}),\qquad
\cre{c}_{R,i}=\frac{1}{\sqrt{2}} (\cre{a}_{i}+\cre{b}_{i}).
\label{basisconversion}
\end{equation}
Given that the new mode functions are sufficiently localized within
each well, the parameters of $\op{H}_{\rm eff}$ for $s=1$ can be
written as
\begin{equation}
 V_a=E-J,\quad V_b=E+J,\quad J_{\alpha,\beta}=J/2,\quad U_{\alpha,\beta}=U/2,
 \label{paramVsJUE}
\end{equation}
where $J=\int{\rm d}x \,\varphi_{R,i}^*  \op{h}_{0,x}
\varphi_{L,i+1}$, $E=\int{\rm d}x \,\varphi_{L,i}^*  \op{h}_{0,x}
\varphi_{L,i}$ and $U=g \int{\rm d}x \,|\varphi_{R,i}|^4$. Notice
that the parameters shown in figure~\ref{parameters} are consistent
with these equations. Expressing the Hamiltonian \eqref{finalh} in
terms of the operators $\cre{c}_{\alpha,i}$ ($\alpha=L,R$) and using
the parameters \ref{paramVsJUE} we find that
\begin{equation}
H_{\rm eff}'=  -J\sum_{i'=1}^{2M-1} \cre{c}_{i'}\des{c}_{i'+1} +{\rm
h.c.} + E\sum_{i'=1}^{2M} \cre{c}_{i'}\des{c}_{i'}+
\frac{U}{2}\sum_{i'=1}^{2M} \cre{c}_{i'} \cre{c}_{i'}
\des{c}_{i'}\des{c}_{i'},\label{BHM}
\end{equation}
where
\begin{equation}
\cre{c}_{i'}=\cases{\cre{c}_{L,\frac{i'+1}{2}}& if $i$ odd,\\
\cre{c}_{R,\frac{i'}{2}}& if $i$ even.\\}
\end{equation}

Therefore, as expected, $\op{H}_{\rm eff}(s=1)$ is equivalent to the
one-band Bose-Hubbard model (BHM)~\cite{jaksch1} with $2M$ sites.

The mode functions keep their symmetry with the two peaks moving
towards the center of the cell when $s$ is decreased. When $s
\approx 0$ is reached $w_{a,i}(x,s)$ and $w_{b,i}(x,s)$ are
equivalent to the Wannier functions of the first and the second
Bloch band, respectively. In this limit we obtain a standard two
band Bose-Hubbard model for $M$ sites.

\section{Time-scale for the preparation of the quantum register}

In this section we present numerical as well as analytical results
characterizing the ground state properties of the system and the
time-scale necessary to initialize the quantum register.

\subsection{Numerical results}\label{numerics}
The numerical calculations have been carried out using both the
exact matrix representation of the Hamiltonian $\op{H}_{\rm eff}$
and the Time-Evolving Block Decimation (TEBD) algorithm (see \ref{tebdalgo} for details).

We have evaluated the ground state and dynamical properties of our
system for two different values of $g$ corresponding to different interaction regimes. These values have been chosen such that in the large
lattice limit (with filling factor $n=1$) the ground state is a
Mott-insulator state for $g_1$ and $g_2$ with 
$J_{aa}/U_{aa}<0.3$~\cite{Kuhner2000}. In the small lattice limit,
$g_1$ and $g_2$ produce ground states corresponding to a
strongly interacting Tonks-Girardeau (TG) gas ($U/J=215$) and a superfluid ($U/J=7$),
respectively.

\subsubsection{Ground states properties}
\label{grounds}

\image{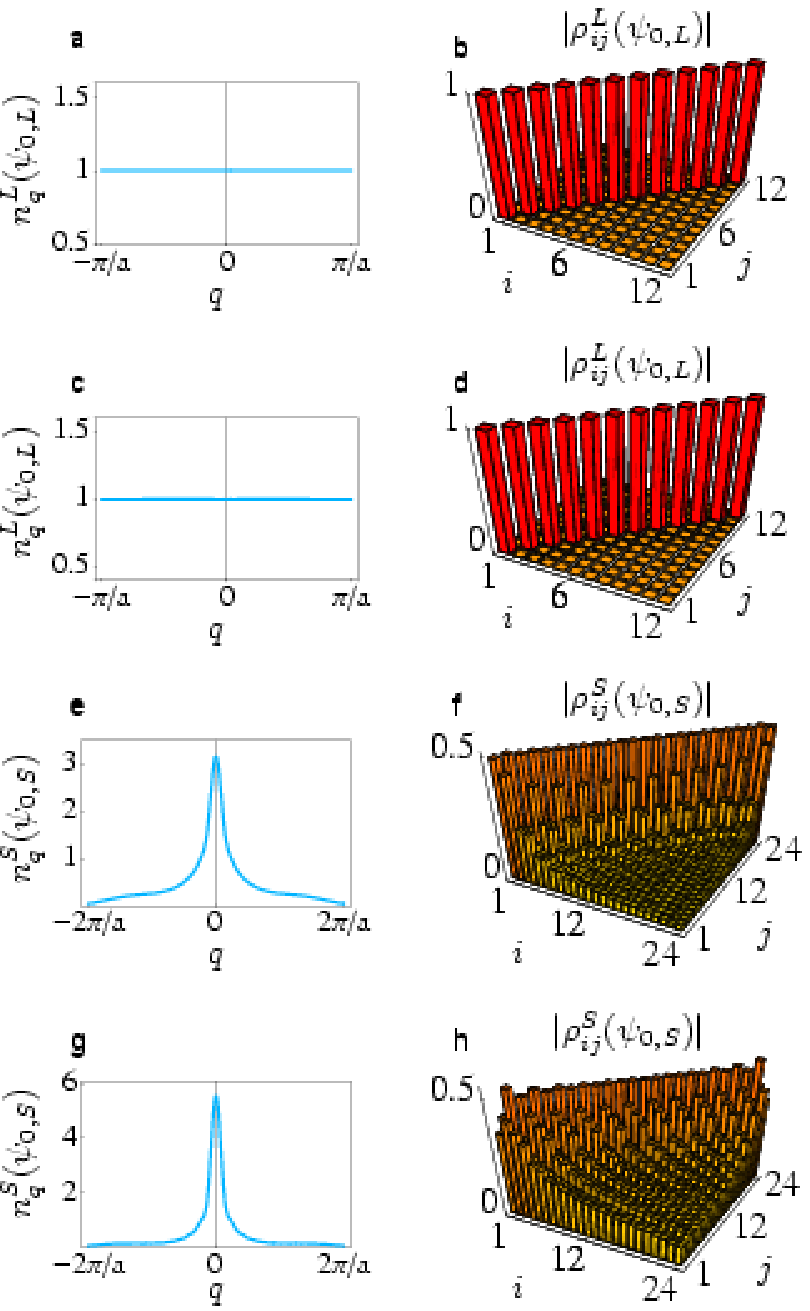}{10cm}{ Ground state one-particle density matrix and
quasi-momentum distribution for $M=12$ lattice sites and the
parameters of figure~\ref{parameters}. The figures (a--d) correspond to the large lattice limit and the figures (e--h) to the small lattice limit. The quasi-momentum distribution (a), (e) and one-particle density matrix (b), (f) are for $g=g_1$. The quasi-momentum distribution (c), (g) and one-particle density matrix (d), (h) are for $g=g_2$. }{gs}

In the large lattice limit, the ground state $\ket{\psi_{0,L}}$ of
the system is populated exclusively by particles in the lowest Bloch
band, i.e.~$a$--mode particles. Therefore, in this limit, we only
consider the one-particle density matrix given by
\begin{equation}
\rho_{ij}^L(\psi)=\bra{\psi} \cre{a}_i \des{a}_j \ket{\psi},
\end{equation}
where $\ket{\psi}$ is the state of the system. The quasi-momentum
distribution of particles in the $a$-mode is given
by~\cite{Roth2003}
\begin{equation}
n_q^L(\psi)=\frac{1}{M} \sum_{i,j=1}^M\,\expim{q a (i-j)}\,\rho^L_{ij}(\psi).\label{qdistrib}
\end{equation}
As expected, in this limit and for commensurate filling $n=1$ the
ground state is a Mott-insulator for both values of $g$ (see
figures~\ref{gs}a--d). The quasi-momentum distributions
show that in this limit particles are uniformly distributed in the
first band (see figures~\ref{gs}a and~\ref{gs}c).

In the small lattice limit, the one-particle density matrix is given
by
\begin{equation}
\rho_{i'j'}^S(\psi)=\bra{\psi} \cre{c}_{i'} \des{c}_{j'} \ket{\psi},
\end{equation}
where the operators $ \cre{c}_{i'}$ are constructed from the
$\cre{a}_i$ and $ \cre{b}_i$ operators using the transformations
\eqref{basisconversion}. The quasi-momentum distribution is given by
\begin{equation}
n_q^S(\psi)=\frac{1}{2M} \sum_{i',j'=1}^{2M}\,\expim{q \frac{a}{2}
(i'-j')}\,\rho^S_{i'j'}(\psi). \label{qdistrib}
\end{equation}
In this limit (with filling factor $n=1/2$), the characteristics of
the ground state $\ket{\psi_{0,S}}$ depend on the value of $g$. For
$g=g_1$, the ratio $U/J = 215$ and the ground state's
correlations as well as the quasi-momentum distributions (see
figures~\ref{gs}e--f) are characteristic of a TG gas (see
e.g.~\cite{Paredes2004} and references therein). The value $g=g_2$
yields the ratio $U/J=7$ and the system exhibits the
behaviour of a superfluid (see
figures~\ref{gs}g--h), with particles occupying mainly the $q=0$
momentum state.  Notice that one-dimensional systems described by
the BHM with a fixed filling factor $n=1$ cross the MI-SF phase
boundary at the critical point  $(J/U)_c \approx
0.3$~\cite{Kuhner2000}. Thus, in our case  the superfluid behavior
of the system in the small lattice limit is due to the fractional
filling of the lattice.

\subsubsection{Simulation of the dynamics}
\label{dsim}

\image{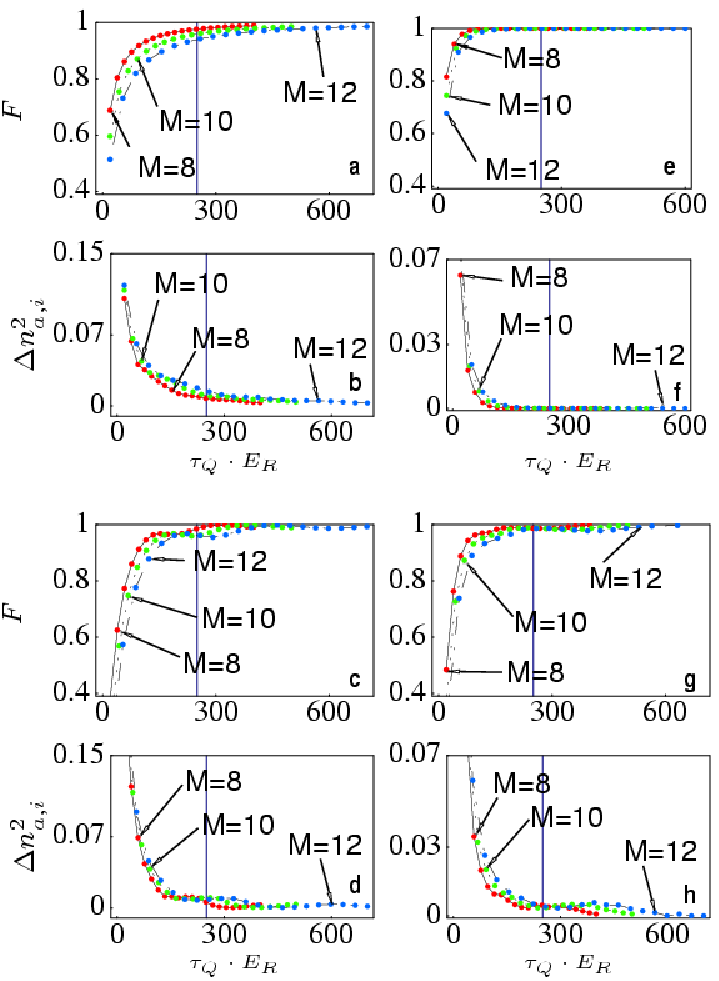}{14cm}{Dynamical simulation of $\op{H}_{\rm
eff}$ using the parameters shown in figure~\ref{parameters} for two
different values of $g$. The ramping strategies $s_{\rm gap}$ and
$s_{\rm man}$ are shown in figure \ref{ramps}a. Simulation results
for the fidelity and particle-number fluctuations using: (a--b) a
linear ramp and $g=g_1$; (c--d) a linear ramp and $g=g_2$; (e--f)
the ramp $s_{\rm gap}$ and $g=g_1$; (g--h) the ramp $s_{\rm man}$
and $g=g_2$.  The vertical lines indicate the value of the particle
tunneling time ($t_{\rm tun}\sim 1/J_{aa}$) in the large
lattice limit.}{fid}

Starting from a system with half-filling and a lattice profile
$s=1$, we investigate the time-scale required to
obtain a nearly perfect MI state---or quantum register---with
filling factor $n=1$ by ramping the lattice profile down to $s=0$.
The quality of the register is determined by the fidelity
\begin{equation}
F=\left| \braket{\psi}{\psi_{0,L}}\right|^2,
\end{equation}
defined as the overlap between the state of the system $\ket{\psi}$
at the end of the ramping process and the ground state in the large
lattice limit $\ket{\psi_{0,L}}$. Furthermore, we calculate the
fluctuations of the number of particles in the $a$-mode at site $i$
\begin{equation}
\Delta n_{a,i}=\sqrt{\langle \left(\op{n}^a_i\right)^2\rangle -\left< \op{n}^a_i\right>^2},
\end{equation}
where $\left< \circ \right>=\bra{\psi}\circ \ket{\psi}$. Since
particle-number fluctuations are suppressed in a MI state, non-zero
fluctuations indicate the presence of excitations, such as double
occupancies or particles in the second band, in the final state.

We test different ramps by simulating the system dynamics between
$t_{\rm i}=0$ to $t_{\rm f}=\tau_Q$ where $\tau_Q$ is the ramping
time (the time required to complete the ramping process from $s=1$ to
$s=0$). Here, each ramp $\sigma$ corresponds to a function
$s=s_\sigma(t)$ varying from $s_\sigma(0)=1$ to
$s_\sigma(\tau_Q)=0$.

For linear ramping we use $s_{\rm lin}(t)=(\tau_Q-t/E_R)/\tau_Q$. The fidelity and particle-number
fluctuations obtained using this strategy for different quench times
and $g=g_1$ and $g=g_2$ are shown in figures~\ref{fid}a--d. The linear ramp is shown in figure~\ref{ramps}a.

Another ramping strategy we use consists of adapting the velocity of
the ramp proportionally to the energy gap between the ground and the
first excited state. This ramp is denoted by $s_{\rm gap}(t)$. We
evaluate the ramp function $s_{\rm gap}(t)$ numerically (see
\ref{rampgap}) for a system with $M=4$ sites and $V_0=30
E_R$. The ramp $s_{\rm gap}(t)$ for $g=g_1$ is shown in figure~\ref{ramps}a. We expect this ramping strategy to
be more efficient than the linear one, since accelerating the ramp
when the gap is large while slowing it down when the gap is small
should suppress transitions of particles to excited levels. Our
numerical calculations have shown that when $g=g_1$, the utilization
of this ramp does indeed reduce the quench time needed to obtain a
nearly perfect fidelity to a half of the tunneling time in the large
lattice limit (see figure~\ref{fid}e--f). For $g=g_2$, we find that
compared to $s_{\rm lin}(t)$, this strategy only marginally improves
the fidelity.

\image{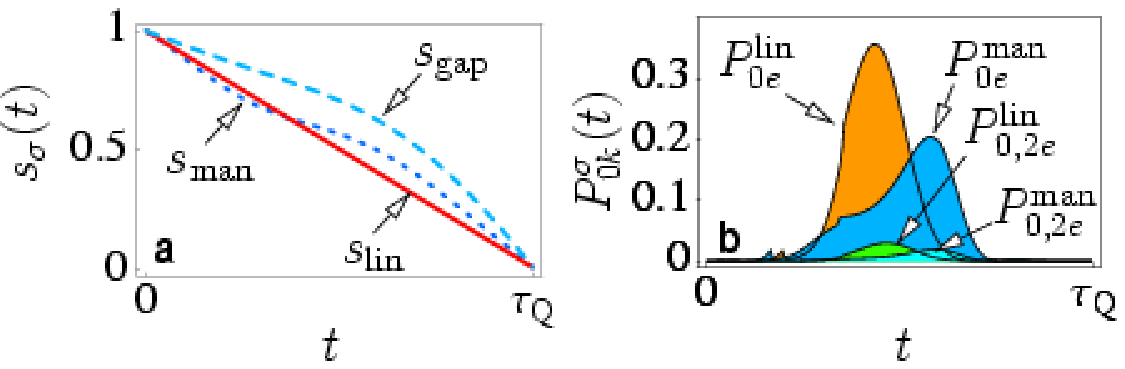}{12cm}{(a) Different ramps used in our numerical
calculations. (b) Transition probability between the ground and the
first (labeled by $e$) and second (labeled by $2e$) excited states
as a function of time for the ramps $s_{\rm lin}$ and $s_{\rm man}$.
The transition probabilities have been calculated via the exact
diagonalization of $\op{H}_{\rm eff}$ for $M=4$ and a quench time of
$\tau_Q=20/E_R$.}{ramps}

In order to reduce the time required to obtain a given fidelity for
$g=g_2$, a more sophisticated ramping strategy is needed. In the
following, we provide a simple method to estimate the efficiency of
different ramps without running a complete dynamical simulation of
the system. The transition probability between the ground and some
excited state $\ket{k}$ at time $t$ for a ramp $\sigma$  is
approximately given by~\cite{Schiff1968}
\begin{equation}
P_{0k}^\sigma(t) \approx  \frac{2}{\omega_{0k}^2}\left| \bra{0}\frac{\rm d}{{\rm d}t}\ket{k}\right|^2 \left[ 1-\cos(\omega_{0k}t)\right],
\label{transitionProb}
\end{equation}
where $\ket{k}=\ket{k(\sigma,t)}$ is the $k$-th instantaneous
eigenstate\footnote{Note that neglecting non-adiabatic changes of
the GWFs does not correspond to $P_{0k}^\sigma(t)=0$ at all times.}
of $\op{H}_{\rm eff}(s_\sigma(t))$ and $\omega_{0k}$ is the
transition frequency between the ground state and $\ket{k}$ .
Therefore, the assessment of the efficiency of a ramp can be
done by evaluating the functional
\begin{equation}
A(\sigma,\tau_Q)=\frac{1}{\tau_Q}\sum_k \int_0^{\tau_Q}{\rm d}t\, P_{0k}^\sigma(t),
\end{equation}
where the index $k$ runs over all the values associated with levels
connected to the ground state. The functional $A(\sigma,\tau_Q)$
corresponds to the average transition probability per unit time for
a given strategy $\sigma$ and quench time $\tau_Q$. We calculate the
value of the functional $A(\sigma,\tau_Q)$ numerically via exact
diagonalization of $\op{H}_{\rm eff}$ for a small system. This
method allows to optimize ramps by minimizing the value of
$A(\sigma,\tau_Q)$. The optimized ramp for a small system is
then used in the simulation of larger systems. For instance, the strategy $s_{\rm man}(t)$ shown
in figure~\ref{ramps}a was designed and optimized manually using
this method. For $g=g_2$, we find that $A({\rm lin},\tau_Q)/A({\rm
man},\tau_Q)\approx 2.3$ for $\tau_Q=20 /E_R$ (see
figure~\ref{ramps}b), thus showing the better efficiency of the
strategy $s_{\rm man}(t)$ compared to $s_{\rm lin}(t)$. As shown in
figure~\ref{fid}, dynamical simulations of the system with $g=g_2$
confirm that this strategy reduces the time required to obtain a
given fidelity.

For systems initially in the superfluid regime ($g=g_2$), the
fidelity curves exhibit small oscillations (see figure~\ref{fid}e).
These can be understood from time-dependent perturbation theory as
oscillations occurring when some of the frequencies involved in the
Fourier decomposition of the perturbation Hamiltonian enter into
resonance with system frequencies. This is expected since
superfluids have a dense spectrum at low energies and are therefore
likely to enter into resonance with one of the frequencies of the
perturbation Hamiltonian~\cite{Clark2006}. Hence, the amplitude of
the oscillations in the fidelity curve associated with a ramping
strategy  $s_1(t)$ should be smaller than those associated with a
ramping strategy $s_2(t)$ if $A(1,\tau_Q)< A(2,\tau_Q)$ for all
$\tau_Q$.  This is what is observed from our numerical simulations
(see figures~\ref{fid}f-h).

\image{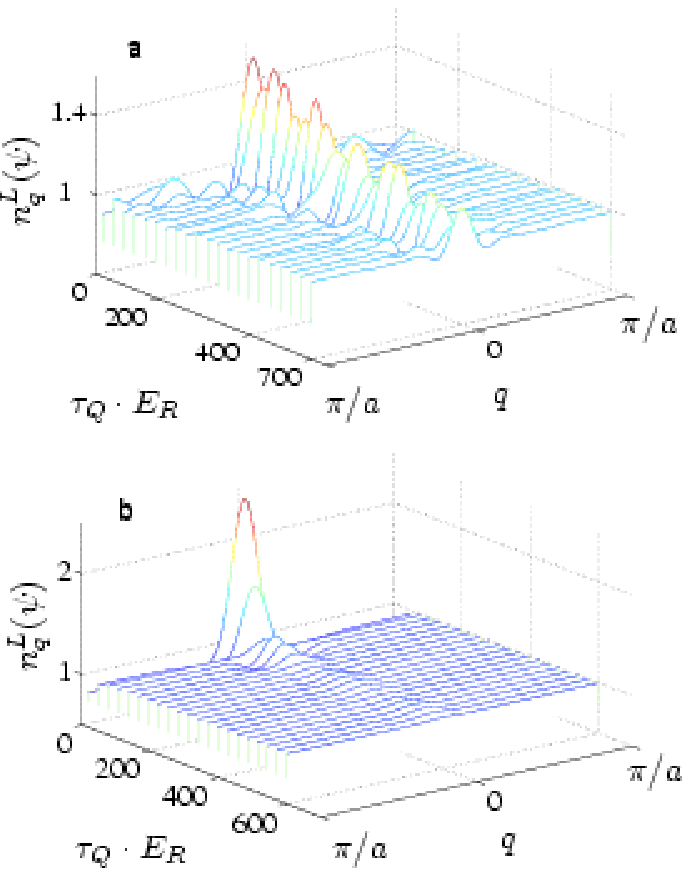}{8cm}{The quasi-momentum distributions at
the end of a ramp for a system of $M=12$ sites using the parameters
shown in figure~\ref{parameters} for $g=g_1$. (a) $s_{\rm lin}$; (b)
$s_{\rm gap}$ .}{simulqdis}

The quasi-momentum distribution of the particles in the $a$--mode at
the end of the different ramping processes for a system with $g=g_1$ are shown in
figure~\ref{simulqdis}. For the linear ramp, the quasi-momentum distribution of particles shown in figure~\ref{simulqdis}a does not correspond to that of a MI state for the quench times considered. In contrast, for the ramp $s_{\rm gap}$ the quasi-momentum distribution becomes approximately flat for quench times of $\tau_Q>200/E_R$. Even for the fastest ramps, we find that the
occupation of the $b$--mode is less than 2\% of the total number of
particles. Thus, the experimental measure of the register fidelity
can be made by comparing the quasi-momentum distribution of
particles in the final state with, e.g. that shown in
figure~\ref{gs}c.

\subsubsection{Discussion of the numerical results}
\label{ts}

In the BHM with $U/J\ll 1$, the tunneling time $1/J$ determines the
adiabatic time-scale of the system. However, as soon as many-body
interactions are sufficiently large, this time-scale often becomes
very non-adiabatic~\cite{Denschlag2002}. The main observation that
can be drawn from the numerical calculations presented in the last
section is that by preparing the system in a TG state ($g=g_1$), and
using an efficient ramping strategy, it is possible to initialize a
very deep MI state on a time scale equivalent to half the tunneling
time in the large lattice limit (see figure~\ref{fid}b). The time required to initialize a MI state with unit filling as well as a TG state with half filling is approximately ten times the tunneling time of the final system~\cite{Blakie2004, Denschlag2002,Griessner2006,Clark2004}. Since the tunneling time in the large lattice limit is two orders of magnitude larger than in the small lattice limit, the total time required to initialize a MI state using our procedure is an order of magnitude faster than the direct quantum freezing method. In this estimation we assume that the initial BEC has zero temperature, i.e., we do not take the effect of defects present in the initial state into account.

The experimental realization of initial states with $g=g_1$ and $g=g_2$ can
be achieved using Feshbach resonances. For a magnetic Feshbach resonance
fluctuations in the magnetic field result in fluctuations of the size of
the gap between the ground and the first excited state which will affect
the performance of our scheme. For instance, in the case $g=g_1$ magnetic
field fluctuations of $10\, \rm  mG$ will change the gap by approximately 0.5\%
for $^{85}\rm Rb$ or $^{133}\rm Cs$ atoms~\cite{Kohler2006,Weber2003}. We assume adiabatic evolution of the
system and thus these fluctuations will have negligible repercussions on
the fidelity of the final state. To realize the superfluid regime with
$g=g_2$ very stable magnetic fields are required. Magnetic field
fluctuations of e.g. $1\, \rm mG$ will lead to gap fluctuations of approximately
1\% in $^{23}\rm Na$ and $^{85}\rm Rb$. We finally remark that our scheme could also be
used without employing a Feshbach resonance. This case corresponds to an
intermediate value of $g$ between $g_1$ and $g_2$. While a detailed analysis
of the intermediate regime is beyond the scope of the present work we do
not expect qualitative differences compared to the interaction strengths
considered here.

\subsection{Analytical results}\label{analytics}

In this section we derive an approximate expression for the quench
time required to obtain a given fidelity in the case of a linear
ramp. 

The energy spectrum of systems in the TG and superfluid regime is
gapless~\footnote{In finite size systems, the spectrum is not gapless, only very dense.}(see e.g.~\cite{Pethick2002}), while in
the Mott-insulating regime, the gap between the ground and first
excited state is proportional to the on-site interaction
energy~\cite{Clark2006}. Since the relaxation time $\tau(t)$ of the
system---the time required by the system to adjust to a change of parameters at time $t$---is inversely proportional to the
gap between the ground and the first excited state, the relaxation time in the small lattice limit is large,
while it is small and finite  in the large lattice limit. This
observation suggests that the {\it adiabatic-impulse} (AI)
assumption from Kibble-Zurek (KZ) theory can be used to evaluate the
adiabaticity of a ramp with respect to the quench
time~\cite{Damski2005,Damski2005a,Cucchietti2006}.

The AI approximation is based on the following considerations: (i)
When the gap between the ground and the first excited state is
large, the relaxation time of the system is short and thus a system
starting its evolution in the ground state remains in
the ground state, i.e. its evolution is {\it adiabatic}. (ii)
When the gap between the ground and the first excited state is
small, the system's relaxation time is large and the system no
longer adapts to changes of the Hamiltonian's parameters and its
state becomes effectively frozen. The system is then in the {\it
impulse regime}.  The instant $\hat{t}$ at which the system passes
from the impulse to the adiabatic regime, and inversely, is defined
by the equation~\cite{Damski2005a,Damski2005,Zurek1985}
\begin{equation}\tau(\hat{t})=\alpha \hat{t},\label{KZeq}
\end{equation}
where $\alpha=\mathcal{O}(1)$ is a constant. Note that $\hat{t}$ is
a time and not an operator. In the AI approximation, the time-evolution of the system is either adiabatic or
impulse. Thus, the {\it density of defects}  ${\mathscr D}$, which
corresponds to the density of excitations caused by a change of
parameters which drive the system from the impulse to the adiabatic
regime can be approximated by~\cite{Damski2005a}
\begin{equation}
{\mathscr D}\simeq \left| \braket{\Psi_e(\hat{t})}{\Psi_g(0)}\right|^2,
\label{densdefect}
\end{equation}
where $\ket{\Psi_g(0)}$ and $\ket{\Psi_e(\hat t)}$ are the ground and first excited states at the initial time $t=0$ and at time $t=\hat t$, respectively.  Hence, without solving the time-dependent Schr\"{o}dinger equation it is possible to make predictions for the density of defects \eqref{densdefect} resulting from a given dynamical process.

In order to apply the KZ theory to our problem, we develop an effective model describing the system dynamics. A similar model was recently examined by Cucchietti {\it et al.}~\cite{Cucchietti2006}. We find from numerical calculations (see figure~\ref{ramps}b) that most of the excitations created in the system are caused by transitions from the ground to the first excited state.  Furthermore, we examined the form of the eigenvectors of the Hamiltonian \eqref{finalh} in both limits for different system sizes via exact diagonalization. This revealed that both the ground state and the first accessible excited state can be approximated by an expansion in only two basis states. The elements of this reduced basis set are given by
\begin{equation}
\ket{1}= \bigotimes_{i=1}^M\, \ket{1}_i,
\end{equation}
\begin{eqnarray}
&&\ket{2}=( \ket{2;0;1;\cdots;1}+\ket{2;1;0;1;\cdots;1} + \ket{2;1;1;0;1;\cdots;1}\nonumber\\
&& +\ket{1;2;1;0;1;\cdots;1}+\cdots)/ \sqrt{M(M-1)},\label{firstx}
\end{eqnarray}
where, e.g. $\ket{2;0;1;\cdots;1}=\ket{2}_1\otimes \ket{0}_2\otimes\ket{1}_3\otimes\cdots\otimes\ket{1}_M$ with $\ket{n}_i=(1/\sqrt{n!})(\cre{a}_i)^n \ket{{\rm vac}}$. The basis state $\ket{2}$ corresponds to a superposition of all possible states of a system of $M$ particles in the $a$-modes with $(M-2)$ singly occupied sites, one doubly occupied site, and one empty site. In the limit $M\rightarrow\infty$, the matrix representation $\op{H}_R$ of Hamiltonian \eqref{finalh} in the reduced basis $\left\{ \ket{1},\ket{2}\right\}$  reads, up to a constant energy $V_a$
\begin{equation}
\op{H}_R =\left( \begin{array}{cc} 0 & \sqrt{2} \,J_{aa}(t)\\ \sqrt{2} \,J_{aa}(t)  & U_{aa}(t) \end{array}\right).\label{HR}
\end{equation}
The instantaneous eigenstates of equation \eqref{HR} associated with the energies of the ground and first excited levels are given by $\ket{g(t)}=-\sin (\theta (t)/2) \ket{1}+ \cos (\theta (t)/2)\ket{2}$ and  $\ket{e(t)}=\cos (\theta (t)/2) \ket{1}+ \sin(\theta (t)/2)\ket{2}$, respectively, with $\cos (\theta (t))=-U_{aa}(t)/[U_{aa}(t)^2+8J_{aa}(t)^2]^{\frac{1}{2}}$, $\theta \in [0,\pi]$. 
Furthermore, we approximate the parameters $J_{aa}$ and $U_{aa}$ as linear functions of time

\begin{equation}
\eqalign{J_{aa}(t)=\Delta J_{aa} (\tau_Q-t)/\tau_Q,\cr
U_{aa}(t)=U_{{\rm init}}+\Delta U_{aa} t/\tau_Q, }
\end{equation}
where $\Delta J_{aa}=|J_{aa}(s=1)-J_{aa}(s=0)|$ , $\Delta U_{aa}=|U_{\rm
init}-U_{aa}(s=0)|$ and $U_{{\rm init}}={\rm min}[U_{aa}]$.

We further simplify the Hamiltonian $\op{H}_R$ by replacing the
hopping term $J_{aa}(t)$ by its time average. Setting
$J_{aa}(t)=\bar{J}_{aa}$ with
$\bar{J}_{aa}=(1/\tau_Q)\int_0^{\tau_Q} {\rm d}t\,J_{aa}(t)$ and
rescaling the time as $t \rightarrow t' -(U_{{\rm init}}
\tau_Q/\Delta U_{aa})$ yields the transformation
\begin{equation}
\op{H}_R \rightarrow \op{H}_R=\left( \begin{array}{cc} 0 & \sqrt{2} \,\bar{J}_{aa}\\ \sqrt{2} \,\bar{J}_{aa}  & \Delta U_{aa} t'/\tau_Q \end{array}\right),\label{HRfinal}
\end{equation}
which turns $\op{H}_R$ into the Landau-Zener form $\op{H}_R=\hat At'
+\hat B$, where $\op{A}$ and $\op{B}$ are Hermitian matrices and
$\op{A}$ is diagonal~\cite{Zener1932,Sinitsyn2004,Rubbmark1981}.
The energy spectrum of the Hamiltonian \eqref{HRfinal} reproduces
approximately the features of the spectrum of $\op{H}_{\rm eff}$
except that the gap is overestimated in the small lattice limit.

Starting at $t=0$ in the small lattice limit where the system is
impulse, we use the AI approximation to derive the density of
defects at the end of a linear ramp which drives the system to the large
lattice limit, where the system is adiabatic. Defining the
relaxation time as the inverse of the energy gap $1/\Delta E$
between the levels of $\op{H}_R$,  with $\Delta E= [(\Delta
U_{aa}t/\tau_Q)^2+ (2 \sqrt{2} \bar{J}_{aa} )^2]^{\frac{1}{2}}$,
equation \eqref{KZeq} can be solved analytically and the instant
$\hat{t}$ at which our system exits the impulse regime is given by
\begin{equation}
\hat{t}=\sqrt{\frac{\tau_Q}{\Delta U_{aa}}} \sqrt{
\sqrt{\frac{1}{\alpha}+(\eta \tau_Q)^2}- \eta \tau_Q},
\end{equation}
where $\eta = 4  \bar{J}_{aa}^2/ \Delta U_{aa}$. We evaluate the
density of defects $\mathscr D$ using equation \eqref{densdefect}
with $\ket{\Psi_g(t_i)}=\ket{g(0)}$ and
$\ket{\Psi_e(\hat{t})}=\ket{e(\hat{t})}$. In order to simplify the
expression of $\mathscr D$, we have set $\theta(0)=\pi/2$, which is
a greater value than the one we would obtain using the real
parameters of the system. This has no significant physical
consequences in our case as it is actually equivalent to considering
a smaller energy gap in the small lattice limit. The density of
defects is then given by
\begin{equation}
{\mathscr D}=\frac{1}{2}(1-\sin \hat{\theta})\label{densdefectfinal},
\end{equation}
where $\hat{\theta} = \theta(\hat{t})$. Inverting equation
\eqref{densdefectfinal}, we obtain
\begin{equation}
\tau_Q({\mathscr D})= \frac{1}{\alpha} \frac{\Delta U_{aa}}{16 \bar{J}_{aa}^2} \frac{(1-2{\mathscr D})^2}{\sqrt{{\mathscr D}(1-{\mathscr D})}},
\label{tauq}
\end{equation}
which gives an approximate analytical expression of the quench time
$\tau_Q$ required to obtain a density of defects $\mathscr D$.

\image{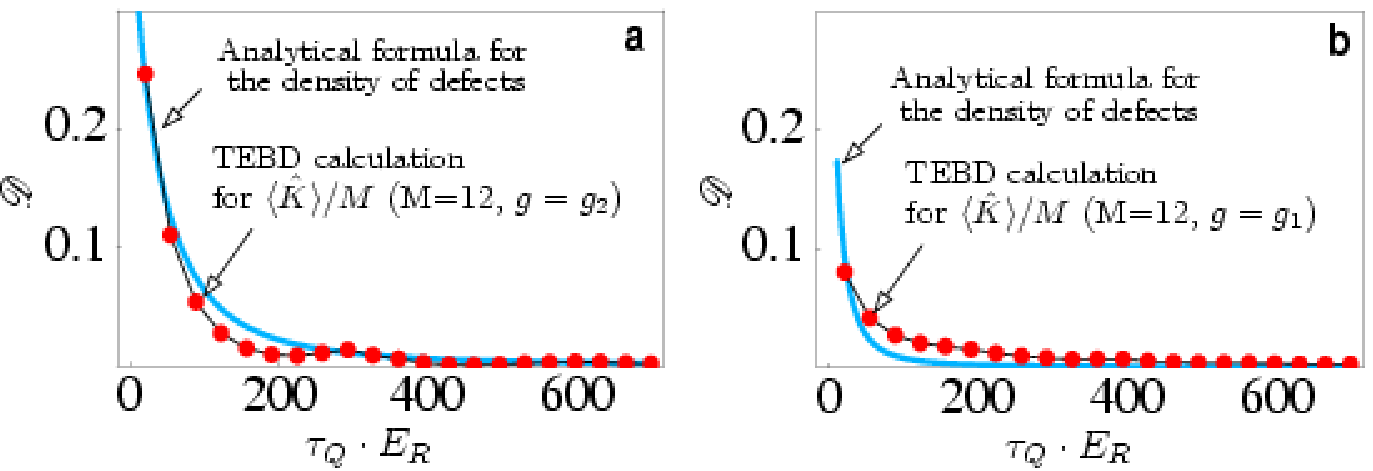}{13cm}{Comparison between the analytical
formula for the density of defects (solid line) and the results of the TEBD calculations for the
density of kinks (dots) $\mathscr D=\langle \op{K} \rangle /M$, for a system
of $M=12$ sites with (a) $g=g_2$ and $\alpha=\sqrt{2}$ and (b) $g=g_1$ and $\alpha=154$. Using the parameters shown in
figure~\ref{parameters}, we have  $U_{\rm init}/g_2=6.8$, $\Delta
U_{aa}/g_2=16.4$, $\bar{J}_{aa}/E_R=0.027$.}{defect}

In our system, defects correspond mainly to doubly occupied sites.
Thus, the number of defects is approximately measured by the
operator
\begin{equation}
\op{K}=\sum_{i=1}^M\, \des{n}^a_i ( \des{n}^a_i - 1).
\end{equation}
Hence, the density of defects is given by ${\mathscr D}=\langle
\op{K} \rangle /M=\langle (\des{n}_i^a)^2 \rangle- \langle
\des{n}_i^a \rangle$. Numerical calculations show that the density
of defects ${\mathscr D}$ has the same scaling behaviour as $\Delta
n_{a,i}^2$. A comparison between equation \eqref{densdefectfinal} and the
numerical values of the density of defects in the large lattice limit
for different values of the ramping time $\tau_Q$ is shown in figure~\ref{defect}. For $M=12$ particles and $g=g_2$, we find that the analytical formula for ${\mathscr D}$ fits the numerical data well for $\alpha=1.41$. For $g=g_1$ the fit is less accurate. This is expected since for this value of $g$, the system has a less distinct separation between the adiabatic and impulse regime than for $g=g_2$. For the number of particles we have been able to simulate, the fit improves as we increase the number of particles for both values of $g$. In addition to this, we see from equation
\eqref{tauq} that the time required to initialize a register with a
given fidelity---and thus the adiabatic time for small $\mathscr
D$---scales with the ratio $\Delta U_{aa}/\bar{J}_{aa}^2$.

\section{Conclusion}\label{conclusion}

We have shown that the dynamics of an optical lattice whose
periodicity is doubled via superlattice potentials is very well
described by a two-mode Hubbard-like Hamiltonian. The parameters of
this Hamiltonian have been evaluated in the tight binding
approximation using optimally localized GWFs. The doubling of the
period removes half of the lattice sites and doubles the filling
factor. We have shown that this doubling can be used for the fast
initialization of a quantum register. By starting from a half filled
lattice in the small lattice limit filled by either a TG ($g=g_1$)
gas or a superfluid ($g=g_2$), a commensurate MI state corresponding
to an atomic quantum register can obtained on timescales shorter
than those achieved by direct quantum freezing of a superfluid with
same lattice spacing. Furthermore, we derived an analytical
expression for the density of defects as a function of the quench
time for linear ramping of the superlattice. We found that the time
required to achieve a given density of defects is proportional to
the ratio $\Delta U_{aa}/\bar{J}_{aa}^2$.

Our numerical calculations of ground state properties suggest that
doubling the lattice period drives the system through a quantum
phase transition for large lattices $M\rightarrow \infty$. The
eventual abrupt change in the ground state properties might be
observable by time-of-flight measurements. An investigation of
whether such a quantum phase transition indeed exists is beyond the
scope of the current work but will be investigated in future work.

In this work we concentrated on the transition from filling factor
$n=1/2$ to $n=1$. We finally note that the idea developed in this paper may be extended by
considering lattices with an initial filling factor of $n=1/2^\ell$
(where $\ell$ is an integer). Subsequently removing every second
barrier will create a lattice with period $2a$ and filling factor
$1/2^{\ell-1}$. This procedure could be repeated $\ell$ times
providing a lattice with filling factor $n=1$ and large lattice
spacing $2^{\ell}a$.

\section{Acknowledgments}
This work was supported by the EU through the STREP project OLAQUI
and a Marie Curie Intra-European Fellowship within the 6th European
Community Framework Programme. The research was also supported by
the EPSRC (UK) through the QIP IRC (GR/S82716/01) and project
EP/C51933/01. DJ thanks the Beijing International Center for
Mathematical Research at Peking University for hospitality while
carrying out parts of this work.

\appendix
\section*{Appendix}
\section{Definition and localization properties of GWFs}\label{gwf}
In the TB limit, the effective single particle Hamiltonian of our
system in momentum space (in the basis $\ket{\alpha}_q=[(1/\sqrt{M})\sum_{i=1}^M \, \expi{q a i} \cre{\alpha}_i]\ket{0}$ with $\alpha=a,b$) can be written as 
\begin{equation}
\op{H}_{0,q}=\sum_{\mu=0,1} \, \op{\varepsilon}(\mu)\,\expim{a \mu q}.
\end{equation}   
The elements of the matrices $\op{\varepsilon}(\mu)$  are given by~\cite{Sporkmann1994}
\begin{equation}
\varepsilon_{\alpha \beta}(\mu)=\bra{w_{\alpha,i}}\op{h}_{0,x}\ket{w_{\beta,i+\mu}}.\label{eigenH}
\end{equation}
For $\mu=0$ and $\mu=1$, the elements
of the matrices $\op{\varepsilon}(\mu)$ correspond to the local site
energy and hopping matrix elements between neighbouring sites,
respectively. For the TB approximation to be accurate, we need the eigenvalues
$E_{\alpha,q}$ of $\op{H}_{0,q}$ to reproduce very closely the band
structure of the exact single particle Hamiltonian $\op{h}_{0,x}$
for all values of the lattice profile $s$. If we were to use WFs as
mode functions $w_{\alpha,i}$, the matrices $\op{\varepsilon}$ would
be diagonal~\cite{Schnell2002} and the dispersion relations of the
two Bloch bands sinusoidal. In order to obtain a more accurate
description, we use GWFs as mode functions. The definition of GWFs
is given by
\begin{equation}
w_{\alpha,i}(x,s)=\sqrt{\frac{a}{2\pi}}\, \int_{-\frac{\pi}{a}}^{\frac{\pi}{a}}{\rm d}q\,\expim{q R_i}\,\tilde{\Psi}_{\alpha,q}(x,s),\label{gwannier}
\end{equation}
where $\tilde{\Psi}_{\alpha,q}=\sum_{\beta=a,b}U^{(q)*}_{\beta,\alpha}\,
\Psi_{\beta,q}$ is called the {\it generalized-Bloch
orbital} with $\Psi_{\alpha,q}$ the Bloch function associated with the band $\alpha$ and $R_i$ is the center of site $i$~\cite{Bross1971,Marzari1997}. The rows of the $2\times 2$
matrix $U^{(q)}$ contain the (real) normalized
eigenvectors of $\op{H}_{0,q}$ associated with the eigenvalues
$E_{\alpha,q}$, that is $\sum_{\mu=a,b} (\op{H}_{0,q})_{n,\mu}
U^{(q)}_{\alpha,\mu}=E_{\alpha,q} U^{(k)}_{\alpha,n}
$~\cite{Sporkmann1994}. Inserting GWFs in equation~\eqref{eigenH},
we recover the elements $\varepsilon_{\alpha \beta}(\mu)$ and, hence, GWFs correspond to the mode functions associated with the effective Hamiltonian $\op{H}_{0,q}$~\cite{Sporkmann1994}. Notice
that the definition of GWFs reduces to that of WFs for
$U^{(q)}=\mathbbm{1}$.

\subsection*{Localization properties of GWFs}

Given a valid set of GWFs, another equally valid set of GWFs can be
obtained by applying the following transformation on the $U^{(q)}$
matrices
\begin{equation}
U^{(q)}\,\rightarrow\,\left( \begin{array}{cc} \expi{\phi_a(q)} &0 \\ 0 &  \expi{\phi_b(q)} \end{array}\right)\,U^{(q)},\label{gaugefreedom}
\end{equation}
where $\phi_\alpha(q)$  are (real) functions of $q$ which can be
chosen freely as long as they do not introduce discontinuities in
the generalized Bloch function~\cite{Blount1962}. The gauge
transformation \eqref{gaugefreedom} is equivalent to re-phasing each
Bloch function  as
$\Psi_{\alpha,q}\,\rightarrow\,\expi{\phi_\alpha(q)}\Psi_{\alpha,q}$.
Notice that gauge transformations do not affect the value of the
parameters $V_\alpha$ and $J_{\alpha,\beta}$ calculated using the
relations \eqref{paramEgen} and  \eqref{paramJgen}, respectively,
but they alter the localization properties---the spread---of the
GWFs. Following the convention suggested by
Blount~\cite{Blount1962}, we set the phase functions
$\phi_\alpha(q)$ in a manner that leads to maximally localized GWFs.
That is, we choose the phase functions such that the resulting GWFs
minimize the spread functional
\begin{equation}
\Omega= \sum_{\alpha}\left<x^2 \right>_\alpha -\left<x \right>_\alpha^2,\label{spreadop}
\end{equation}
where in our case $\alpha=a,b$ while
$\left<x^2 \right>_\alpha= \bra{w_{\alpha,i}}x^2  \ket{w_{\alpha,i}} $ and $\left<x
\right>_\alpha= \bra{w_{\alpha,i}}x  \ket{w_{\alpha,i}} $ correspond to the center
of a GWF and its second moment, respectively.

Expanding $\tilde{u}_{\alpha,q}(x,s)=\expim{q x} \tilde{\Psi}_{\alpha,q}(x,s)$ into plane waves yields
\begin{equation}
\tilde{u}_{\alpha,q}(x,s) = \sum_{j}\,  G_{\alpha,j}(q,s)\,\expi{K_j
x},
\end{equation}
where $K_j=2 \pi j/L$ with $L=M a$. Invoking the translational
symmetry of the lattice and the convolution
theorem~\cite{Blount1962}, the value of the functional $\Omega$ is
minimized when the expansion coefficients $G_{\alpha,j}(q,s)$ are
chosen real, which is always possible when the lattice possesses
mirror symmetry ~\cite{Sporkmann1994}. This is equivalent to the
choice of purely real GWFs for even generalized-Bloch functions and
purely imaginary ones for odd generalized-Bloch functions. Notice
that this conclusion is in agreement with a conjecture of Marzari
{\it et al.}~\cite{Marzari1997} on the real nature (up to a global
phase) of maximally localized WFs.

We have numerically evaluated the phases $\phi_\alpha(q)$ using the
algorithm described in~\cite{Marzari1997} for the special case of 1D
WFs. This procedure minimized the functional $\Omega$ in the limit
of very fine sampling of the $q$-space. The effect of this
localization procedure is illustrated in figure~\ref{wannier}a.

\section{Parameters of the single-particle Hamiltonian}\label{paramEq}
We diagonalize $\op{H}_0$ in momentum space for periodic boundary
conditions. Using the Fourier transformations
$\des{a}_i=(1/\sqrt{M})\sum_{q} \, \expi{q a i} \op{a}_q$ and
$\des{b}_i=(1/\sqrt{M}) \sum_{q} \, \expi{q a i} \op{b}_q$,
$\op{H}_0$ becomes block diagonal
\begin{equation}
\op{H}_0=\sum_{q}\,\op{H}_{0,q},
\end{equation}
where $q= 2 \pi \nu /M a$, $\nu=1\dots M$. In the basis
$\left\{\ket{a}_q,\ket{b}_q\right\}$, the operator $\op{H}_{0,q}$
reads
\begin{equation}
\op{H}_{0,q} =\left( \begin{array}{cc} V_a -2 J_{aa}\,\cos qa & -{\rm i}2 J_{ab}\, \sin qa \\  {\rm i}2 J_{ab} \,\sin qa & V_b+2J_{bb}\,\cos qa \end{array}\right),\label{htq}
\end{equation}
with $\ket{a}_q=\cre{a}_q\ket{0}$ and $\ket{b}_q=\cre{b}_q\ket{0}$. For simplicity, the explicit dependence of the parameters on the lattice profile has been dropped.

Due to the periodicity of the lattice, the eigenvalues of $\op{H}_{0,q}$ exhibit a band structure. By choosing the points $q=0,\,\pi/a,\,\pi/2a$ in the Brillouin zone, we derive and solve a set of equations for the parameters $V_\alpha$ and $J_{\alpha,\beta}$ as functions of the eigenvalues $E_{\alpha,q}$ of $\op{H}_{0,q}$
\begin{equation}
\eqalign{
J_{aa} = \frac{E_{a,\pi}-E_{a,0}}{4} \quad, \quad J_{bb} = \frac{E_{b,0}-E_{b,\pi}}{4},\cr
V_a =\frac{E_{a,0}-E_{a,\pi}}{2}\quad,\quad V_b=\frac{E_{b,0}-E_{b,\pi}}{2},\cr
J_{ab} =\frac{1}{4} \left[(E_{b,\frac{\pi}{2}}-E_{a,\frac{\pi}{2}})^2 -(V_a-V_b)^2\right]^{\frac{1}{2}}.}
\label{paramVsband}
\end{equation}
 For the eigenvalues of $\op{H}_{0,q}$ to reproduce the band structure of the exact single-particle Hamiltonian along the $x$--direction $\op{h}_{0,x}$, we evaluate the parameters $V_\alpha$ and $J_{\alpha,\beta}$ using the eigenvalues $\varepsilon_{\alpha,q}$ of $\op{h}_{0,x}$ obtained via exact numerical diagonalization for the same points in the Brillouin zone (see figure~\ref{bandstructure}). The numerical values of the parameters obtained via this procedure are shown in figure~\ref{parameters}a.

The accuracy of the TB approximation can be tested by evaluating the standard deviation between the exact and approximated band structure. That is, taking $N_q$ different points $q_i$ on each band, we define the standard deviation between the exact and approximate band structure for a given lattice profile by $\sigma^2_s=(1/2N_q)\,\sum_{i=1}^{N_q} ( \Delta \varepsilon_{a,q_i}^2+\Delta \varepsilon_{b,q_i}^2)$, with $\Delta \varepsilon_{\alpha,q_i}=\varepsilon_{\alpha,q_i}-E_{\alpha,q_i}$. Averaging over $N_s$ different lattice profiles $s_i$, we obtain $\left<\sigma\right>_s=[(1/N_s) \sum_{i=1}^{N_s}\sigma^2_{s_i}]^{\frac{1}{2}}=3.4\times10^{-2} E_R$ for a lattice depth of $V_0= 10\, E_R$. This excellent agreement improves further as we increase the value of $V_0$, and hence fully justifies the tight-binding approximation.

\section{Dynamical and ground state calculations using the TEBD algorithm}\label{tebdalgo}
The TEBD algorithm is based on directly manipulating a matrix
product representation of the many-body wave function. Here, we shall
briefly describe the key aspects of this algorithm and refer the
reader to some of the recent literature ~\cite{Vidal2003,Vidal2004,Vidal2006,Daley2004a} for more detail.

An arbitrary state of a 1D quantum lattice system composed of $M$
sites can be written as
\begin{equation}
\ket{\psi}=\sum_{j_1=1}^{\mathbbm{d}}\cdots\sum_{j_M=1}^{\mathbbm{d}}c_{
j_1
\cdots j_M}\ket{j_1,\dots,j_M} \nonumber,
\end{equation}
where $c_{j_1 \cdots j_M}$ is a set of $\mathbbm{d}^M$ complex
amplitudes and $\ket{j_m}$ is a basis spanning the local
$\mathbbm{d}$-dimensional Hilbert space of site $m$. Within
time-dependent DMRG the amplitudes $c_{j_1 \cdots j_M}$ are
constructed from a product of tensors
\begin{equation}
\label{expansion} c_{j_1 j_2 \cdots
j_M}=\sum_{\{\alpha\}=1}^{\{\chi\}}\Gamma^{[1]j_1}_{\alpha_1}\lambda^{[1
]}_{\alpha_1}
\Gamma^{[2]j_2}_{\alpha_1\alpha_2}\lambda^{[2]}_{\alpha_2}\cdots\Gamma^{
[M]j_M}_{\alpha_{M-1}},
\end{equation}
where $\{\alpha\}=\{\alpha_1,\cdots,\alpha_{M-1}\}$, $\{\chi\} =
\{\chi_1,\cdots,\chi_{M-1}\}$ and with $\Gamma$ and $\lambda$
tensors chosen to be constructed from the set of $M-1$ Schmidt
decompositions for contiguous partitions of the system.
Specifically, the elements of $\lambda^{[m]}_{\alpha}$ are taken
to be Schmidt coefficients of the bipartite splitting after site
$m$ in $\ket{\psi}=\sum_{\alpha=1}^{\chi_m}\lambda^{[m]}_{\alpha}
\ket{L^{m}_{\alpha}}\ket{R^{m}_{\alpha}}$ with Schmidt rank
$\chi_m$. The Schmidt states $\ket{L^{m}_\alpha}$ and
$\ket{R^{m}_\alpha}$ spanning the left $\{1,\cdots,m\}$ and right
$\{m+1,\cdots,M\}$ subsystems of sites respectively are then
specified by the corresponding sums remaining in
equation~\eqref{expansion}.

The usefulness of this representation is based on the observation
that for 1D systems with a Hamiltonian composed of nearest
neighbour terms the groundstate and low-lying excited states have
Schmidt coefficients $\lambda^{[m]}_{\alpha}$ which rapidly decay
with $\alpha$ when arranged in descending order. Consequently,
rather than allowing the Schmidt ranks $\chi_m$ to grow to their
maximum permissible value a much smaller fixed upper-limit $\chi$
can be imposed truncating the representation while still providing
a near unit overlap with the exact state $\ket{\psi}$. Fixing the
Schmidt ranks results in the number of parameters scaling as
$O(\mathbbm{d}\chi^2M)$ and so curtails the possible exponential
growth with $M$ seen for general coefficients $c_{j_1 \cdots
j_M}$.

The matrix product representation also permits the efficient
update of the state after the action of a unitary operator on any
two neighboring lattice sites. This proceeds by modifying the
$\Gamma$ tensors associated to the sites and the $\lambda$ tensor
linking them and requiring a number of operations which scales as
$O(\mathbbm{d}^4 \chi^3)$. The resulting tensors are then
systematically truncated back to a maximum rank of $\chi$.

Dynamical simulations can be performed by decomposing the time
evolution operator $\exp(-iH\delta t)$, for small time step $\delta
t$, into a sequence of pairwise unitaries via a Suzuki-Trotter
expansion. Given the properties outlined such a calculation is
likely to be accurate for a practical value of $\chi$ if both the
initial state and the states generated by the dynamics remain in the
low-energy manifold of the system. To determine the appropriate
$\chi$ calculations are repeated with increasing values of $\chi$
until the final result converges and are unaffected by further
increases. For practical purposes the convergence is usually
quantified by the robustness of the expectation values calculated.
The accuracy of a calculation is also gauged by the sum of the
discarded Schmidt coefficients at each time step - a quantity which
should necessarily be small - and the deviation of normalization of
the final state from unity which indicates the accumulated effect of
truncation.

Finally, initial states are typically taken to be the groundstate
of the system which are found either by applying the DMRG
procedure or, as in this work, by simulating imaginary time
evolution through the repeated application of $\exp(-H\delta t)$
and subsequent renormalization of the state. In our simulations, we have used $\chi=40$.

\section{The ramp $s_{\rm gap}$}\label{rampgap}
The ramp $s_{\rm gap}$ can be evaluated as follows. The energy
gap between the ground and the first excited state associated with a
given lattice profile $s$ is given by ${\rm gap}(s)= \omega_{0e}$
where $\omega_{0k}=(\varepsilon_{{\rm eff},k}-\varepsilon_{{\rm
eff},0})$ with $\varepsilon_{{\rm eff},\alpha}$ the $\alpha$-th
instantaneous eigenvalue of $\op{H}_{\rm eff}(s)$. The gap is
evaluated numerically via the direct diagonalization of $\op{H}_{\rm
eff}$ for a small number of sites. The function $s_{\rm gap}(t)$ is
defined by the equation ${\rm d}s_{\rm gap}(t)/{\rm d}t={\rm
gap}(s_{\rm gap}(t))/K$ where $K^{-1}=\int_{0}^{\tau_Q}{\rm d}t\,
{\rm gap}(s_{\rm gap}(t))$ is a normalization constant.

\section*{References}
\bibliography{articleNJP}
\end{document}